\documentclass[a4paper,11pt]{article}

\usepackage{jheppub} 

\usepackage{bbm}
\usepackage{graphicx}  
\usepackage{dcolumn}   
\usepackage{bm}        
\usepackage{amssymb}   
\usepackage{braket}
\usepackage{amsmath}
\usepackage{color}  	
\usepackage{slashed}
\usepackage{subfigure}
\usepackage[utf8]{inputenc}
\DeclareGraphicsExtensions{.pdf,.png,.jpg,.eps}
\graphicspath{ {./images/} }

\newcommand{\Tr}{\mathrm{Tr}}

\newcommand{\td}{\mathrm{d}}

\newcommand{\TT}[1]{\mathrm{#1}}

\renewcommand{\v}[1]{\mathbf{#1}}

\renewcommand{\>}{\right\rangle}
\newcommand{\<}{\left\langle}
\newcommand{\lkl}{\left|}
\newcommand{\rkl}{\right|}

\newcommand{\qqqquad}{\qquad\qquad\qquad}
\newcommand{\qqqqquad}{\qquad\qquad\qquad\qquad}

\newcommand{\Nc}{N_{\mathrm{c}}}
\newcommand{\As}{\alpha_{\mathrm{s}}}
\usepackage{breakurl}

\hypersetup{urlcolor=black}

\title{Comments on a new `full colour' parton shower}

\preprint{\begin{flushright}
    MAN/HEP/2020/001 \\
    UWTHPH-2020-9\\
  MCnet-20-10\end{flushright}
}    
	
\author[a,b]{Jack Holguin,}
\author[a,b]{Jeffrey R. Forshaw,}
\author[b,c]{Simon Pl\"{a}tzer.}

\affiliation[a]{Consortium for Fundamental Physics, School of Physics \& Astronomy, \\
	University of Manchester, Manchester M13 9PL, United Kingdom}
\emailAdd{jack.holguin@manchester.ac.uk}
\emailAdd{jeffrey.forshaw@manchester.ac.uk}
\emailAdd{simon.plaetzer@univie.ac.at}
\affiliation[b]{
  Erwin Schr\"{o}dinger Int. Institute for Mathematics and Physics, \\
  University of Vienna, 1090 Wien, Austria}
\affiliation[c]{Particle Physics, Faculty of Physics, \\
	University of Vienna, 1090 Wien, Austria}
\date{\today}

\abstract{A new parton shower algorithm has been presented with the
  claim of providing soft-gluon resummation at `full colour'
  \cite{Hoeche:2020nsx}. In this paper we show that the algorithm does
  not succeed in this goal.  We show that full colour accuracy
  requires the Sudakov factors to be defined at amplitude level and
  that the simple parton-shower unitarity argument employed in
  \cite{Hoeche:2020nsx} is not sufficient.}

\begin{document} 

\maketitle
\flushbottom

\section{Introduction}

Over recent years much attention has been devoted to the development
of parton showers with `full colour' evolution
\cite{Nagy:2012bt,Platzer:2012np,SoftEvolutionAlgorithm,Platzer:2018pmd,Forshaw:2019ver,Nagy:2019pjp}. The
study of these has multiple motivations: most importantly, reducing
theoretical uncertainties in parton showers will be crucial for
precision phenomenology at future colliders. Currently, parton showers
provide some of the largest sources of uncertainty in experimental
analyses, e.g. \cite{Azzi:2019yne}. There has also been a growth in
interest towards developing tools for the formal resummation of
observables sensitive to the complexity of the non-abelian structure
of the strong interaction, specifically observables with non-global or
super-leading logarithms
\cite{SuperleadingLogs,Banfi:2006gy,factorisationBreaking,Hagiwara:2015bia,Caron-Huot:2015bja,Becher:2016mmh}. These
will play an important role in advancing parton shower algorithms. In
this context, a widely available `full colour' parton shower would be
a powerful tool.

In this letter we comment on the formalism for resumming complex
colour structures employed recently in \cite{Hoeche:2020nsx}. A
similar approach was previously put forward by one of the present
authors and collaborators \cite{Platzer:2012np,Platzer:2018pmd}. The
authors of \cite{Hoeche:2020nsx} describe their formalism as being
capable of producing ``numerical resummation at full color in the
strongly ordered soft gluon limit.'' We will examine this claim in
what follows.

Let us be clear on what we mean by leading and sub-leading colour. A
general observable can be written
\begin{align}
\Sigma(L) = \sum^{\infty}_{n=0} (\Nc\As)^{n} \sum^{n+1}_{m = 0} C_{n,m}(L) \ ,
\end{align}
where $L$ is some large logarithm. The coefficients $C_{n,m}$ can be
expanded:
\begin{align}
C_{n,m} = \underbrace{C^{(0)}_{n,m}}_{\TT{LC}_{\Sigma}} +  \underbrace{\frac{1}{\Nc}C^{(1)}_{n,m}
 }_{\TT{NLC}_{\Sigma}} + \underbrace{\frac{1}{\Nc^2}C^{(2)}_{n,m}}_{\TT{NNLC}_{\Sigma}} + ... \,  \label{eq:colorexp}
\end{align}
and a `full colour' shower should be able to compute all of the
$C_{n,m}^{(i)}$ at a stated logarithmic accuracy.\footnote{Or in a
  specified kinematic limit, e.g. the strongly-ordered soft gluon
  limit.} We will show that the formalism of \cite{Hoeche:2020nsx}
generally fails to compute the NNLC$_{\Sigma}$ terms, even in the
strongly-ordered soft gluon approximation. Note also that, for many
observables, the NLC$_\Sigma$ term vanishes, so that the dominant
sub-leading colour corrections occur at NNLC$_\Sigma$.  It is also
important to appreciate that the colour expansion defined in
Eq.~\eqref{eq:colorexp} is very weak in its ambition. Just as in the
case of logarithmic resummation, more ambitious would be to perform a
resummation of towers of enhanced corrections. In which case an
expansion of the form of Eq.~\eqref{eq:colorexp} would be
exponentiated.

\section{Summary of the new `full colour' parton shower}

We will briefly summarize the algorithm advocated in
\cite{Hoeche:2020nsx} and we largely follow their notation. The
amplitude for an $n$-parton hard process is $\lkl {M}_{n} \>$ and
$\lkl m_{n+k} \>$ is the amplitude after dressing with $k$ soft
gluons. Real emissions are accounted for recursively according to
\begin{align}
\< m_{n+k} | m_{n+k} \> = \< m_{n+k-1} \rkl \v{\Gamma}_{n+k-1}(\v{1}) \lkl m_{n+k-1} \> = \< M_{n} \rkl \v{\Gamma}_{n} ( ... \v{\Gamma}_{n+k-2}(\v{\Gamma}_{n+k-1}(\v{1}))...) \lkl M_{n} \>, \label{eqn:reals}
\end{align}
where
\begin{align}
\v{\Gamma}_{n}(\v{\Gamma}) = - \sum^{n}_{\stackrel{i,j=1}{i \neq j}} \v{T}_{i} \, \v{\Gamma} \, \v{T}_{j}  \; \omega_{ij}, \qquad \omega_{ij} = \frac{s_{ij}}{s_{iq}s_{qj}}
\end{align}
and $s_{ij}=2p_i\cdot p_j$ in terms of the momenta of the partons $i$
and $j$. The radiation pattern for a single emission, $q$, is then
determined by 
\begin{align}
\frac{\td \sigma_{n+k+1}}{\sigma_{n+k}} = \td \Phi_{+1} 8\pi \As \frac{\< m_{n+k} \rkl \v{\Gamma}_{n+k}(\v{1}) \lkl m_{n+k} \>}{\< m_{n+k} | m_{n+k} \>},
\end{align}
where $\td \Phi_{+1}$ is a phase-space measure and parametrises the
momentum map from a state of $n+k$ partons to a state of $n+k+1$
partons. Its details are not needed for our discussion. Virtual
corrections are encoded via a no-emission probability, i.e. via a
typical parton-shower cross-section-level Sudakov factor, defined
though unitarity as
\begin{align}
\int^{t}_{t'} \frac{\td \kappa^{2}_{ij}}{\sigma_{n+k}} \int \frac{\td \sigma_{n+k+1}}{\td \kappa^{2}_{ij}} \Pi(\kappa^{2}_{ij}, t) = 1 - \Pi(t', t),
\end{align}
where $\kappa^{2}_{ij} = \omega^{-1}_{ij}$ plays the role of the
ordering variable.  This equation has the solution
\begin{align}
\Pi^{(k)}(t', t) =  \prod^{n+k}_{\stackrel{i,j=1}{i \neq j}} \Pi_{ij}(t', t),
\end{align}
where 
\begin{align}
\Pi_{ij}(t', t) = \exp \left(-\int^{t}_{t'} \frac{\td \kappa^{2}_{ij}}{\kappa^{2}_{ij}} \int\frac{8\pi\td \Phi_{+1}}{\td \kappa^{2}_{ij}} \; \As \frac{\< m_{n+k} \rkl \v{T}_{i} \, \v{T}_{j} \lkl m_{n+k} \>}{\< m_{n+k} | m_{n+k} \>}\right)\ ,
\end{align}
is the no-emission probability for a single dipole $(i,j)$. The
overall no-emission probability dresses the real emission matrix
elements defined in Eq.~\eqref{eqn:reals} according to
\begin{align}
\< m_{n+k} ; t| m_{n+k} ; t \> = \Pi^{(k)}(t, t_{k}) ... \Pi^{(1)}(t_{2}, t_{1})\Pi^{(0)}(t_{1}, Q^{2})\< m_{n+k}| m_{n+k}\>,
\end{align}
where $t_{i}$ is the ordering variable associated with the $i$th
emission and $Q^{2}$ is the hard scale.

\section{The problem with Sudakovs}

In this section we show that defining Sudakov factors through
cross-section-level unitarity gives rise to two compounding errors in
colour. The first error is in the computation of loops, the second is
in the computation of the interplay between loops and real
emissions. These errors make the inclusion of Coulomb terms
impossible, since they always appear as a pure (abelian) phase in the
amplitude. Firstly, we address the computation of loops (resummed into
Sudakov factors). The role of Sudakov factors in full-colour evolution
of amplitudes has been extensively studied
\cite{Nagy:2008eq,Platzer:2013fha,Nagy:2015hwa,Caron-Huot:2015bja,Becher:2016mmh,SoftEvolutionAlgorithm,Forshaw:2019ver,Nagy:2019pjp}. Ignoring
Coulomb terms (including them only makes matters more complicated),
Sudakov factors\footnote{The argument of the Sudakov exponent is the
  real part of the one-loop cusp anomalous dimension
  \cite{Caron-Huot:2015bja,Becher:2016mmh}. Depending on the choice of
  ordering variable, path ordering should be implied. See Section 2 of
  \cite{Forshaw:2019ver} for more details.} should dress a general
amplitude as
\begin{align}
&\< m_{n+k};t' | m_{n+k};t' \>  \nonumber \\ 
&= \< m_{n+k} ; t\rkl e^{-\int^{t}_{t'} \td \kappa^{2} \int\frac{4\pi\td \Phi_{+1}}{\td \kappa^{2}} \; \As\v{\Gamma}_{n+k}(\v{1})}e^{-\int^{t}_{t'} \td \kappa^{2} \int\frac{4\pi\td \Phi_{+1}}{\td \kappa^{2}} \; \As\v{\Gamma}^{\dagger}_{n+k}(\v{1})} \lkl m_{n+k} ; t\>, \nonumber \\ 
&= \frac{\< m_{n+k} ; t \rkl e^{-\int^{t}_{t'} \td \kappa^{2} \int\frac{8\pi\td \Phi_{+1}}{\td \kappa^{2}} \; \As\v{\Gamma}_{n+k}(\v{1})}\lkl m_{n+k} ; t\>}{\< m_{n+k} ; t| m_{n+k} ; t\>} \< m_{n+k} ; t| m_{n+k} ; t\>,\nonumber \\
& \nonumber \\
& \neq \Pi^{(k)}(t', t)\< m_{n+k} ; t| m_{n+k} ; t \>. \label{eqn:error}
\end{align}
The not equals to sign represents the first error in
\cite{Hoeche:2020nsx}.

We will now attempt to explicate this error and its consequences by
giving it two different interpretations. Firstly, we will show how
this error can be thought of as a straightforward linear algebra
error. Secondly, we will present some fixed-order calculations that
show this error corresponds to miscalculating NNLC$_{\Sigma}$ diagrams
with two or more loops. To begin the linear algebra interpretation,
let us rewrite the pertinent term from Eq.~\eqref{eqn:error} as
\begin{align}
& \frac{\< m_{n+k} ; t\rkl e^{-\int^{t}_{t'} \td \kappa^{2} \int\frac{8\pi\td \Phi_{+1}}{\td \kappa^{2}} \; \As\v{\Gamma}_{n+k}(\v{1})}\lkl m_{n+k} ; t\>}{\< m_{n+k} ; t| m_{n+k} ; t\>} \nonumber \\
&= \frac{\Tr \left(\lkl m_{n+k} ; t\>\< m_{n+k} ; t\rkl e^{\v{V}}\right) }{\Tr \left(\lkl m_{n+k} ; t\>\< m_{n+k} ; t\rkl\right) } \equiv \Tr_{\TT{norm}} \left(e^{\v{V}}\right), \label{eqn:correct}
\end{align}
where $\Tr_{\TT{norm}}$ is a normalised trace, such that
$\Tr_{\TT{norm}} \v{1} = 1 \neq N$ where $N$ is the dimension of the
matrix. In this notation we can write
\begin{align}
\Pi^{(k)}(t', t) = e^{\Tr_{\TT{norm}} \left(\v{V}\right)}.
\end{align}
This definition is the source of the error. Motivated by
cross-section-level arguments of unitarity, it is implicitly assumed
that
\begin{align}
\Tr_{\TT{norm}} \left(e^{\v{V}}\right) = e^{\Tr_{\TT{norm}} \left(\v{V}\right)},
\end{align}
which is wrong.

As a trivial example of how this sort of error could give problems,
consider \mbox{$\Tr \, e^{\v{1}_{N}} = Ne$} whereas $ e^{\Tr \,
  \v{1}_{N}} = e^N$. However, the error from using a normalised trace
is more subtle, since $\Tr_{\TT{norm}} e^{\v{1}_{N}} =
e^{\Tr_{\TT{norm}}\v{1}_{N}} = e$. To see where the actual problem
arises, consider a toy model where $\v{V} = \As\Nc( \v{1} + \Nc^{-1}
\delta \v{V})$ and $\delta\v{V}$ is not diagonal. In this case, the
$\As \Nc \v{1}$ piece plays the role of the leading colour part of the
Sudakov and $\As \delta\v{V}$ the sub-leading colour part. The result
is that
\begin{align}
\Tr_{\TT{norm}} \left(e^{\v{V}}\right) = e^{\Tr_{\TT{norm}} \left(\v{V}\right)} +  \sum_{n\geq 2} \mathcal{O}\left(\As^{n} \Nc^{n-2} (\Tr_{\TT{norm}} \delta \v{V}^{2} - (\Tr_{\TT{norm}} \delta \v{V})^{2})\right).
\end{align}
The important difference arises because $(\Tr_{\TT{norm}} \delta
\v{V})^{n} \neq \Tr_{\TT{norm}} (\delta \v{V}^{n})$ for $n\geq
2$. From this argument it is clear that errors will occur, starting
with the computation of NNLC$_{\Sigma}$.

Now let us now give a physical interpretation of the error by
expanding Eq.~\eqref{eqn:correct} to $\mathcal{O}(\As^{2})$. The
$\mathcal{O}(\As^{2})$ term corresponds to dressing a general hard
process at fixed order with two strongly ordered soft loops. The
correct amplitude is
\begin{align}
&\sum^{n}_{\stackrel{i,j=1}{i \neq j}} \int^{t}_{t'} \td \kappa^{2}_{ij} \int\frac{8\pi\td \Phi_{+1}}{\td \kappa^{2}_{ij}} \; \As \, \sum^{n}_{\stackrel{k,l=1}{k \neq l}} \int^{t}_{\kappa^{2}_{ij}} \td \kappa^{2}_{kl} \int\frac{8\pi\td \Phi_{+1}}{\td \kappa^{2}_{kl}} \; \As \nonumber \\ 
& \qqqqquad \times\Tr_{\TT{norm}} \left(\v{T}_{i} \cdot \v{T}_{j} \; \v{T}_{k}  \cdot \v{T}_{l}\right) \< m_{n+k} ; t| m_{n+k} ; t\>.
\end{align}
Now, we can expand $\Pi^{(k)}(t', t)\< m_{n+k} ; t| m_{n+k} ; t \>$ to
the same order. We find
\begin{align}
&\frac{1}{2}\sum^{n}_{\stackrel{i,j=1}{i \neq j}} \int^{t}_{t'} \td \kappa^{2}_{ij} \int\frac{8\pi\td \Phi_{+1}}{\td \kappa^{2}_{ij}} \; \As \, \Tr_{\TT{norm}} \left(\v{T}_{i} \cdot \v{T}_{j} \right) \sum^{n}_{\stackrel{k,l=1}{k \neq l}} \int^{t}_{t'} \td \kappa^{2}_{kl} \int\frac{8\pi\td \Phi_{+1}}{\td \kappa^{2}_{kl}} \; \As \nonumber \\ 
& \qqqqquad \times \Tr_{\TT{norm}} \left( \v{T}_{k} \cdot \v{T}_{l}\right) \< m_{n+k} ; t| m_{n+k} ; t\>. \label{eqn:wrong2loop}
\end{align}
These two expressions are only equal when $n+k \le 3$ because the
colour matrices are then proportional to identity matrices. However,
for multiplicities of coloured partons greater than $3$ they differ by
NNLC$_{\Sigma}$ pieces. This error occurs because writing a matrix
element in the form of Eq.~\eqref{eqn:wrong2loop} implicitly assumes
that $[\v{T}_{i} \cdot \v{T}_{j} , \v{T}_{i} \cdot \v{T}_{k}] \approx
0$, which is only correct up to NLC$_{\Sigma}$ terms. For example,
consider the case of $e^{+}e^{-} \to q \bar{q} g_{1} g_{2}$ (for which
the NLC$_\Sigma$ term is zero). To illustrate the point consider the
limit that both gluons were emitted from the quark. In this limit a
NNLC$_{\Sigma}$ error emerges due to the non-vanishing of
\begin{align}
\As^{2}\Tr_{\TT{norm}} \left(\v{T}_{q} \cdot \v{T}_{g_{1}} \; \v{T}_{g_{1}}  \cdot \v{T}_{g_{2}}\right) - \As^{2}\Tr_{\TT{norm}} \left(\v{T}_{q} \cdot \v{T}_{g_{1}} \right) \Tr_{\TT{norm}} \left( \v{T}_{g_{1}}  \cdot \v{T}_{g_{2}}\right)\nonumber \\
 = \As^{2}\frac{\Nc^6+3 \Nc^4-14 \Nc^2+2}{4 \Nc^2 (\Nc^2-1)^2} = \frac{(\Nc \As)^{2}}{4}\left(\frac{1}{\Nc^{2}} + \frac{5}{\Nc^{4}} + ...\right).
\end{align}
Similar errors arise from other emission topologies. The non-vanishing
commutator is also the reason why Coulomb terms do not cancel and, as
a result, underpins the origin of super-leading logarithms
\cite{SuperleadingLogs}.

The second error compounds the first. Let us now consider the
evolution of an amplitude to a new scale whilst emitting a single
gluon:
\begin{align}
&\< m_{n+k+1};t'' | m_{n+k+1};t'' \>  \nonumber \\ 
&\qqqquad= \int^{t'}_{t''} \td \kappa^{2} \int\frac{8\pi\td \Phi_{+1}}{\td \kappa^{2}} \; \As \< m_{n+k} ; t\rkl e^{-\int^{t}_{t'} \td \kappa^{2} \int\frac{4\pi\td \Phi_{+1}}{\td \kappa^{2}} \; \As\v{\Gamma}_{n+k}(\v{1})} \v{\Gamma}_{n+k}(\v{1})  \nonumber \\ 
& \qqqquad \quad \times e^{-\int^{t}_{t'} \td \kappa^{2} \int\frac{4\pi\td \Phi_{+1}}{\td \kappa^{2}} \; \As\v{\Gamma}^{\dagger}_{n+k}(\v{1})} \lkl m_{n+k} ; t\>.  \label{eqn:seconderror}
\end{align}
In order to recombine the two exponentials into a single Sudakov that
builds $\Pi^{(k)}(t', t)$ one must assume $[\v{\Gamma}_{n+k}(\v{1}),
  e^{\v{V}}] \approx 0$.  For the same reasons as those described
above, this is again a NNLC$_{\Sigma}$ error. Where the previous error
was in the higher order colour of loop diagrams ($\ge$ 2 loops), this
error is in the higher order colour from the interplay between $(\ge
1)$ loops and emissions. Consequently, the algorithm does correctly generate 
real emissions in the absence of any loop corrections. It also correctly generates
one-loop contributions that dress the softest real emission but fails thereafter. 

\section{Conclusions}
QCD colour dynamics beyond leading colour is highly non-trivial and
its correct inclusion generally requires an amplitude-level approach
that goes beyond the simple treatment of virtual corrections presented
in \cite{Hoeche:2020nsx}.

\section*{Acknowledgements}
The authors want to thank the Erwin Schr\"odinger Institute for
support during the period when this and related work has been carried out.  This work
has received funding from the UK Science and Technology Facilities
Council (grant no. ST/P000800/1), the European Union’s Horizon 2020
research and innovation programme as part of the Marie
Skłodowska-Curie Innovative Training Network MCnetITN3 (grant
agreement no. 722104), and in part by the by the COST actions CA16201
``PARTICLEFACE'' and CA16108 ``VBSCAN''. JH thanks the UK Science and
Technology Facilities Council for the award of a studentship.
\bibliographystyle{JHEP} \bibliography{comments}

\end{document}